\documentclass[apsspec,showpacs,twocolumn,groupedaddress,lengthcheck,preprintnumbers]{revtex4}
\usepackage{amsmath}
\usepackage{amssymb}
\usepackage{graphics}
\usepackage[dvips]{graphicx}
\usepackage{graphicx}
\usepackage[usenames]{color}

\def\simge{\mathrel{%
       \rlap{\raise 0.511ex \hbox{$>$}}{\lower 0.511ex \hbox{$\sim$}}}}
\def\simle{\mathrel{
       \rlap{\raise 0.511ex \hbox{$<$}}{\lower 0.511ex \hbox{$\sim$}}}}

\newcommand \beq{\begin{eqnarray}}
\newcommand \eeq{\end{eqnarray}}

\usepackage[colorlinks=true,linktocpage=true,linkcolor=blue,citecolor=blue]{hyperref}

\begin{document}
\title{Evolution of Primordial Neutrino Helicities in Astrophysical
Magnetic Fields and Implications for their Detection}
\author{Gordon Baym and Jen-Chieh Peng}
\affiliation{\mbox{Illinois Center for Advanced Studies of the Universe}\\}
\affiliation{\mbox{Department of Physics, University of Illinois, 1110
  W. Green Street, Urbana, IL 61801} \\
  }

\date{\today}

\begin{abstract}   
  Since decoupling in the early universe in helicity states, primordial neutrinos propagating in astrophysical magnetic fields precess and undergo helicity changes.   In view of various experimental bounds allowing 
a large magnetic moment of neutrinos, we estimate the helicity flipping for relic neutrinos in both cosmic and galactic magnetic fields.  The flipping probability is sensitive both to the neutrino magnetic moment and the structure of the magnetic fields, and is a potential probe of the fields.   As we find, even a magnetic moment well below that suggested by XENON1T could significantly affect relic neutrino helicities and their detection rate via inverse tritium beta decay. 


\vspace{0.1cm}

\end{abstract}

\maketitle

The early universe was bathed with thermal neutrinos which decoupled from matter around 1 s 
after the big bang.  Detection of these neutrinos, e.g., 
through inverse beta decay capture on tritium~\cite{weinberg} in the PTOLEMY 
experiment~\cite{ptolemy,long}, remains a major challenge.   Relic neutrinos carry information about the early 
universe at a much earlier epoch than that of photon decoupling.  In addition, neutrinos propagating through the universe acquire information about the gravitational and magnetic fields they encounter en route to Earth.  We focus here on the evolution of the helicity of primordial neutrinos and implications for their detection rates.

  Neutrinos in the early universe decoupled essentially in chirality
eigenstates at temperatures orders of magnitude larger than 
neutrinos masses, leaving these highly relativistic neutrinos essentially
in helicity eigenstates as well.  Were neutrinos 
to travel freely from decoupling to the present, we would 
expect neutrinos to be left handed, and antineutrinos 
right handed.  However, two effects modify this conclusion.  
   
The first is that as the neutrino trajectory is bent gravitationally 
by density fluctuations in the universe, the deflection of its 
spin vector lags behind that of its momentum vector; gravitational 
fields do not conserve neutrino helicity \cite{duda,silenko,dvornikov,long,grav}. 
Second, neutrinos of finite mass are expected to have a non-zero 
magnetic moment \cite{marciano,benlee,fujikawa,lynn,s-w,bell-Dirac,bell,dolgov,gs}, 
so that propagation in galactic and cosmic magnetic fields \cite{subra,tanmayreview} rotates their spins with respect to their momenta, again allowing neutrinos and antineutrinos to have an amplitude to be flipped in helicity, as first noted in Ref.~\cite{fujikawa}.
Neutrinos are potential probes of cosmic and galactic magnetic fields as well as density fluctuations in the expanding universe. 

  {  We explore here consequences of large neutrino magnetic 
moments on the time evolution of primordial neutrino helicities.   
Helicity modification by slowly varying astrophysical magnetic fields 
occurs via diagonal magnetic moments and is thus limited to Dirac 
neutrinos.   In contrast, both Dirac and Majorana helicities are 
modified by gravitational fields \cite{grav}.
The recent XENON1T report of an excess of low energy electron
events~\cite{xenon1t} has
triggered interest in the possibility that a large 
magnetic moment of solar neutrinos, of order $\sim1.4-2.9 \times 10^{-11} \mu_B \,  (\equiv \mu_{1T})$, where $\mu_B$ is the Bohr magneton, could account for these excess events~\cite{miranda,babu}.  While beyond-the-Standard Model physics. which is required for large magnetic moments, generally favors sizeable moments for 
Majorana rather than Dirac neutrinos~\cite{bell-Dirac,bell,miranda,babu}, the XENON1T data, which does not distinguish diagonal from transition moments, can accommodate both neutrino types.  We do not
assess here the possibility of moments exceeding estimated 
theoretical bounds \cite{bell-Dirac,enqkvist,semikoz2017}.  }

 We find that even a moment several orders of magnitude smaller 
than $\mu_{1T}$ could lead to significant helicity changes as Dirac neutrinos 
propagate through the cosmos, as well as the Milky Way.   As we discuss, detection
rates for primordial neutrinos are sensitive to both 
their helicity structure as well as whether they are Dirac or 
Majorana fermions.   We assume $\hbar=c=1$ throughout.

We first briefly recall some properties of primordial neutrinos from standard 
cosmology.  At temperatures, $T$, small compared with the muon mass but well above 1 MeV, muons and tau are frozen out 
and the only charged leptons present are electrons and positrons; 
neutrinos are held in thermal equilibrium with the ambient plasma 
through neutral and charged current interactions. 
As estimated in Ref.~\cite{dolgov}, $\nu_\tau$ and $\nu_\mu$ freeze 
out at temperature $T_\mu\sim  1.5$ MeV, while $\nu_e$ freeze out 
at $T_e\sim 1.3$ MeV.   However, the temperature differences 
at freezeout do not effect the present temperature, $T_{\nu 0}= 1.945 \pm 0.001$K = $(1.676\pm 0.001)\times 10^{-4}$ eV, of the various neutrino species.

  The observation of neutrino oscillations establishes well
that neutrino 
flavor eigenstates $\alpha$ are linear superpositions of mass 
eigenstates $i$ with PMNS flavor-mass mixing matrix elements $U_{\alpha i}$ \cite{esteban}.  
While neutrinos
decouple in flavor eigenstates, the velocity disperson, $\delta v = \frac12 \Delta m^2 /p^2$, 
among different mass components of momentum $p$ soon separates a given flavor state
into three effectively decoherent wave packets of mass eigenstates \cite{masses}. 
 The distribution of the 
present momenta $p_0$ of primordial neutrinos of each mass state $|\nu_i\rangle$ is 
\beq
  f(p_0) = \frac{1}{e^{p_0/T_{\nu0}}+1},
\label{f0}
\eeq 
independent of the neutrino mass, 
with total number density $n = 3\eta(3)T^3_{\nu0}/2\pi^2  = 56.25 \,\mbox{cm}^{-3}$.

   The magnetic moment of a non-zero mass Dirac neutrino is estimated in the (extended) standard 
model to be \cite{marciano,benlee,fujikawa}
\beq
   \mu_\nu^{\rm SM} \simeq  \frac {3G_F}{4\sqrt2\,\pi^2} m_\nu m_e \mu_B \simeq  
3 \times 10^{-21} m_{-2}  \mu_B;
   \label{numag}
\eeq
$\mu_B$ = 1.40 MHz/gauss is the Bohr magneton, and $m_{-2}$ the neutrino 
mass in units of $10^{-2}$ eV.    Diagonal moments of Majorana neutrinos 
must vanish, although transition moments connecting different mass 
eigenstates are non-zero  \cite{gs}.   Magnetic moments 
could be substantially larger than Eq.~(\ref{numag}) predicts.  
According to the most recent Review of Particle 
Physics~\cite{RPP}, the most sensitive upper bounds for $\mu_\nu$
are given by the GEMMA and Borexino experiments. The GEMMA reactor 
experiment  \cite{gemma} 
gives an upper limit $\mu_\nu < 2.9 \times 10^{-11} \mu_B$, and  
Borexino \cite{borexino} reports upper bounds from solar neutrinos, 
$\mu_{\nu_e} < 2.8\times 10^{-11}\mu_B$. These bounds are comparable 
to the moment $\mu_{1T}$
that could explain the XENON1T low-energy electron-event 
excess, which does not 
distinguish diagonal from transition magnetic moments.

 As
a neutrino with a magnetic moment propagates through magnetic 
fields its spin precesses, see, e.g.,  \cite{semikoz1993}. 
To set the scale, we first neglect relativistic 
effects; then since the neutrino magnetic moment 
vector is $\mu_B \hat S = 2\mu_B \vec S$,  the 
rotation rate of the spin is 
$\omega_s = 2\mu_\nu B$, where $B$ is a characteristic field strength.  
For example, for
$B \sim10^{-12}$G,  of order present intergalactic magnetic fields,
the rotation rate with (\ref{numag}) becomes $\omega_s \sim 8 \times 
10^{-27} m_{-2}\, \rm Hz$.  Over the total age of 
the universe, $t_0 \sim 4.3\times 10^{17}$ s, the spin would 
rotate by a net angle $2 \mu_\nu B t_0 \sim 4\times 10^{-9}m_{-2}$, and more 
generally,
$ \sim 10^{12}(\mu_\nu/\mu_B)(B/10^{-12} \rm G)$.   Owing, however, 
to magnetic fields being considerably larger in the early universe, 
this result underestimates the spin rotation.
Transition moments do not lead to such spin rotation, and thus 
Majorana neutrinos would not be affected \cite{semikoz1997,barranco}.  

  We calculate the neutrino spin $\vec S$ and its rotation in the neutrino rest frame, measuring transverse and longitudinal spin components $S_\perp$ and $S_\parallel$ with respect to the axis of the neutrino ``lab" momentum, where the lab frame is that of the ``fixed stars."   
For rotation from an initial helicity state, for which $S_{\perp} = 0$, by angle $\theta$, one has $|\vec S_{\perp}| /|\vec S|=\sin\theta$.  The helicity changes from $\pm 1 $ to  $\pm \cos\theta $, and the probability of observing the helicity flipped is then $P_f = \sin^2(\theta/2)$; for $\theta\ll 1$, $P_f \simeq  \theta^2/4$.

    The spin precesses in its rest frame according to
\beq
  \frac{d\vec S}{d\tau} = 2\mu_\nu \vec  S \times \vec  B_R,
  \label{dsdtr}
\eeq
where $\tau$ is the neutrino proper time, and $\vec B_R$ is the magnetic field in the rest frame.  
   
  In terms of the lab frame magnetic field and time, $t$, the equations of motion of the rest frame spin are \cite{BMT},
\beq
      \frac{d\vec S_{\perp}}{dt} &=& 2\mu_\nu\left(\vec S_{\parallel} \times \vec B_{\perp} +\frac{1}{\gamma} \vec S_{\perp} 
                     \times \vec  B_{\parallel }\right), \label{slab} \\                   
          \frac{dS_{\parallel}}{dt}   &=&  2\mu_\nu(\vec S \times \vec B)_\parallel,    \label{slabpar}                     
\eeq
since in the absence of an electric field in the lab frame, 
$B_{\parallel R} = B_{\parallel}$, $B_{\perp R} = \gamma B_{\perp}$, and  $d\tau = dt/\gamma$,  where $\gamma = E_\nu/m_\nu$.
We neglect the $\nu_e$-$e$ matter effect~\cite{Voloshin}, important only 
for very dense matter or vanishingly small $\mu_\nu$.

    For small deviations, $|S_\perp| \ll |S|$, from a pure helicity state, the $\vec S_{\perp} 
                     \times \vec B_{\parallel }$ term in Eq.~(\ref{slab}) is negligible; thus  a neutrino of velocity $\vec v$ and helicity $\pm 1$ experiences a cumulative spin rotation
with respect to its momentum,
\beq
   \frac{\vec S_{\perp}}{|\vec S\,|} = \pm 2\mu_\nu \int dt \, \hat v\times \vec B(t).
        \label{spinrotu}
\eeq

   One of the larger magnetic fields a relic neutrino encounters 
en route to local detectors is that of our galaxy, $B_g \sim 10\mu$G.
Galactic fields do not point in a uniform direction, but rather 
change orientation over a coherence length, $\Lambda_g$, of order
kpc \cite{TPB,zweibel,han2004,YRZ}.   The spin orientation undergoes a random walk 
through the changing directions of $\vec B$, reducing
the net rotation by a factor $\sim\sqrt{\ell_g/\Lambda_g}$, where $\ell_g$ 
is the mean crossing distance of the galaxy, of order the galactic 
volume $V_g$ divided by $\sigma_g$, its cross-sectional area.  Thus the mean
square spin rotation of a neutrino passing through a 
galaxy ($g$) is
\beq
    \langle \theta^2\rangle_{g} \simeq \left(2\mu_\nu B_g 
\frac{\Lambda_g}{v}\right)^2  \frac{\ell_g} {\Lambda_g}
    \label{galrot}
\eeq
All quantities (except $\mu_\nu$) nominally depend on the epoch $t$.    The spin rotation is larger for more massive neutrinos since $1/v^2 = 1+m_\nu^2/p^2$, with $p$ the neutrino momentum. 
    
    The spin rotation for non-relativistic neutrinos ($m_\nu \gg p\simeq T_{\nu0}$), evaluated with parameters 
characteristic of the Milky Way, $B_g \sim 10 \,\mu$G, 
$\ell_g\sim$ 16 kpc, $\Lambda_g \sim $ kpc, is 
\beq
   \langle \theta^2\rangle_{\rm MW} &\sim&   
        4\times 10^{29} m_{-2}^2\left(\frac{\Lambda_g}{1\,{\rm kpc}}\right)   
      \left(\frac{B_g}{10 \mu \rm G}\right)^2 \left(\frac{\mu_\nu}{\mu_B}\right)^2.
      \label{galrot1}  \nonumber\\
\eeq
A moment $\sim 1.5 \times 10^{-15} \mu_B$,  a factor $10^{-4}$ smaller than what would account for the XENON1T excess, would yield a helicity flip probability $P_f$  of order unity for $m_{-2} (B_g/10 \mu{\rm G}) (\Lambda_g/1 {\rm kpc})^{1/2}$ itself of order unity. 

    Neutrinos propagate past distant galaxies before reaching the
Milky Way.    The effective number of galaxies a neutrino sees per unit 
path length is $\sim n_g \sigma_g$, where $n_g$ is the number density 
of galaxies.    Integrated over the neutrino trajectory from early galaxies to 
now, the effective number, $N_{\rm eff}$, of galaxies a neutrino passes 
through is $\sim n_g \sigma_g R_u \sim n_g V_g(R_u/\ell_g)$, where $R_u$ is the present radius of the universe.   Since  $n_gV_g \sim 10^{-6}$ 
and $R_u/\ell_g \sim 10^6$, a neutrino would pass through $N_{\rm eff}$ of 
order unity before reaching the Milky Way.    The cumulative rotation of a neutrino prior to reaching 
our galaxy is comparable to the spin rotation 
it would undergo within the Milky Way.

   We now estimate the net rotation a relic neutrino experiences 
from cosmic magnetic fields in the expanding universe, from decoupling to now.   
We work in the metric $ds^2 = -a(u)^2(du^2-d\vec x\,^2)$,
where $\vec x$ are the co-moving spatial coordinates,  and $a(u)$ is the increasing scale factor of the universe (with $a=1$ at present); the conformal time $u$ is related to the coordinate time by $dt = a(u)du$.  Over the evolution of the universe from decoupling,  where $a(t_d)\equiv a_d \sim 10^{-10}$, to now the cosmic magnetic field 
decreases;  assuming that the field lines move with the overall expansion, flux conservation implies that 
globally $Ba^2$ should remain essentially constant in time.   As with galaxies, the coherence length, $\Lambda$, of the cosmic magnetic field is not well determined, but expected to be
on Mpc scales  \cite{tanmay,neronov,ackermann};  the coherence length 
reduces the net spin rotation by a factor $\sim\sqrt{\Lambda/R_u}$.

  In order of magnitude, the ratio of the helicity flip probability from the present cosmic 
field to that from a galactic field, is:
\beq
  \frac{\langle \theta^2\rangle_{\rm galaxy}}
{\langle\theta^2\rangle_{\rm cosmic}} \sim 
\left(\frac{B_g}{B_u}\right)^2 \frac{\ell_g\Lambda_g}{R_u\Lambda}.
\eeq
The magnetic field ratio is of order of at least microgauss vs. picogauss, 
while the ratio of length scales is of order (kpc)$^2$/(Gpc Mpc)
 $\sim 10^{-9}$, which would indicate a scale of neutrino spin rotation in galaxies up to three orders of magnitude larger 
 than in cosmic magnetic fields.
However, in assessing whether cosmic rotation is competitive with the 
rotation from the galactic magnetic field, it is necessary, in addition to determining better the cosmic and galactic magnetic fields and correlation lengths, to take into account the larger cosmic fields as well as smaller coherence lengths at earlier times.

We turn now to this latter task.  We start from 
the squared rotation in Eq.~(\ref{spinrotu}), written in terms of $u$ for relativistic neutrinos, with $c$ denoting ``cosmic,"
\beq
    \langle \theta^2\rangle_c = 4\mu_\nu^2\ 
\big\langle\big(\int du a(u)\vec B_\perp(u) 
\big)^2\big\rangle_c,
\label{thetasq}
 \eeq
with the expectation value in the cosmic background.    The correlation function of the cosmic magnetic field, in 
an otherwise isotropic background, has the structure
\beq
& \langle B_i(\vec x)B_j(\vec x\,') \rangle &=\\ &&(-\delta_{ij}\nabla^2 + 
\nabla_i\nabla_j)F(r) + \epsilon_{ijk}\nabla_k G(r),  \nonumber
\eeq
where $r=|\vec x-\vec x\,'|$, $F$ is the normal and $G$ the helical 
field \cite{axel} correlation.   The latter does not contribute 
to the spin rotation since $\nabla_z G(r)$ is odd in $\vec x-\vec x\,'$, 
and thus its contribution for transverse spin components, 
$\sim \int du du' \partial_z G(r)$,  vanishes by symmetry. 

  The normal correlation has the Fourier structure \cite{planckmag},
\beq
     \langle B_i(\vec x)B_j(\vec x\,') \rangle = \int 
\frac{d^3k}{(2\pi)^3} \frac{\delta_{ij}-\hat k_i \hat k_j}{2}P_B(k)e^{i\vec k 
\cdot (\vec x - \vec x\,')},
  \nonumber\\
  \label{bcorr}
\eeq
where in another convention for the correlation function \cite{tanmayreview}, 
$P_B(k) = (2\pi)^2E_M(k)/k^2$.    Equation~(\ref{bcorr}) implies that
\beq
   \langle \vec B\,^2\rangle = \int \frac{d^3k}{(2\pi)^3} P_B(k).
   \label{pb}
\eeq
 The schematic structure of $P_B$ is a power law $\sim \alpha k^s$ at 
small $k$ out to a wavevector $k_*$ (called $k_i$ in Ref.~\cite{tanmayreview}) , followed by a sharper 
falloff,  $\sim\beta k^{-q}$ beyond $k_*$, with $q>3$ and 
$\beta = \alpha k_*^{s+q}$.   The sign of $s$ is 
uncertain \cite{kunze,tanmayreview} but infrared convergence of the integral in Eq.~(\ref{kperpint})
requires $s> -2$.   With this approximate form Eq.~(\ref{pb})  implies
$  \alpha \simeq  2\pi^2 (s+3)(q-3) \langle \vec B^2\rangle_c/(s+q)k_*^{s+3}$.

   With Eq.~(\ref{bcorr}) and taking the $z$-axis along the neutrino velocity, Eq.~(\ref{thetasq}) becomes
\beq
    \langle \theta^2\rangle_c &=&4\mu_\nu^2  \int du du' a(u)a(u') 
\times\\
     &&\quad\int \frac{d^3k}{(2\pi)^3} e^{ik_z(u-u')} \frac{1- k_z^2/k^2}{2} P_B(k). \nonumber
    \eeq
Since the scale of $k $ is $\gg 1/u$, the $u$ integrals are vanishingly small except in the 
neighborhood of $k_z=0$, and to a first approximation we set $k_z=0$ in $(1-k_z^2/k^2) P_B(k)$.  Then 
 the $k_z$ integral gives a factor $2\pi\delta(u-u')$, and
\beq
    \hspace{-12pt} \langle \theta^2\rangle_c &\simeq& 
\frac{\mu_\nu^2}{\pi}\int_{u_d}^{u_0} du  a(u)^2\int _0^\infty dk_\perp  k_\perp P_B(k_\perp), 
    \label{kperpint}
\eeq
where $0$ denotes present values, $u_0 = 3t_0$, and $d$ denotes neutrino decoupling.
Here
\beq
 \int _0^\infty dk_\perp  k_\perp P_B(k_\perp) \simeq 2\pi^2\eta \frac{\langle \vec B^2\rangle}{k_*}.
\eeq
 With conservation of flux, $\langle \vec B^2(u)\rangle \simeq B_0^2 /a(u)^4$, and $k_*(u) \sim 2\pi/\Lambda_0 a(u)^{1/2}$ \cite{tanmayreview}.
The factor $\eta = (s+3)(q-3)/(s+2)(q-2)$ is not strongly dependent on the spectral indices, and for simplicity we take $\eta=1/2$ (corresponding to 
$s=2$ and $q= 2+5/3$).  Then 
\beq
       \langle \theta^2\rangle_c &=& \frac12 \mu_\nu^2
         B_0^2 \Lambda_0 \int_{u_d}^{u_0} \frac{du}{a(u)^{3/2}}.
 \label{inter}        
\eeq

    The main contribution to the integral is from the radiation-dominated era, from the time of neutrino decoupling, $u_d$, to the time of matter-radiation equality, $u_{eq}$,  where $a(t_{eq}) \equiv a_{eq} \sim 0.8\times 10^{-4}$.  In this era $a \propto u$, and 
 \beq
 \int_{u_d}^{u_{eq}}\frac{du}{a(u)^{3/2}} \simeq  \frac {2u_0}{a_{eq}^{1/2} a_d^{1/2}}\simeq 
  2\times 10^7 u_0, 
  \eeq
since $u_{eq}$ is related to $u_0$ by $u\propto a^{1/2}$ in the matter-dominated era.
By comparison, in the matter-dominated era,  
\beq
  \int_{u_{eq}}^{u_0}\frac{du}{a(u)^{3/2}}\simeq \frac{u_0}{2a_{eq}},
  \label{mde}
\eeq
a factor $\sqrt{a_d/a_{eq}}/4\sim 10^{-4}$ smaller.

  Altogether
\beq
    &&\hspace{-24pt}   \langle \theta^2\rangle_c \simeq 9\left(\frac{\Lambda_0}{R_u}\right)\,
       \frac{(\mu_\nu t_0 B_0)^2}{a_{eq}^{1/2} a_d^{1/2}}\nonumber\\
       &\simeq& 2\times 10^{27} \left(\frac{\Lambda_0}{1\,{\rm Mpc}}\right)  
\left(\frac{B_0}{10^{-12}\rm G}\right)^2 \left(\frac{\mu_\nu}{\mu_B}\right)^2 , 
    \label{rotresult}
\eeq
independent of the neutrino momentum.  To within uncertainties in magnetic fields, correlation lengths, and neutrino masses, the estimated spin rotation in the cosmos is basically comparable to that in galaxies.

    If the low-energy electron-event excess found in the XENON1T 
experiment \cite{xenon1t} does arise from a neutrino magnetic moment,  
and the neutrino is a Dirac particle, its diagonal moment could lead to a significant spin rotation.     
 A magnetic moment of order $10^{-2}\mu_{1T}$ would still produce a spin rotation in the range 
of detectability.   On the other hand, if the neutrino is a Majorana particle, the excess would occur entirely  
from transition magnetic moments, with no helicity changes from magnetic fields.

  Having described the expected spin rotation of relic neutrinos 
we turn to their detection. The most promising approach 
is to capture neutrinos on beta unstable nuclear targets.  Particularly 
favorable for detecting primordial neutrinos is the inverse tritium beta 
decay (ITBD)\cite{weinberg,long}, $\nu_e + ^3$H$\to ^3$He + e$^-$
the reaction inverse to tritium beta decay,
$^3$H $\to ^3$He + e$^- + \bar \nu_e$.
The ITBD would yield a distinct signature of a mono-energetic peak separated 
from the endpoint of the tritium beta decay by $2m_\nu$. 

   The cross section for capture of a neutrino in mass state $i$ on tritium is  \cite{long}
\beq
 \sigma^h_i(p,p_e)
 &&\\ && \hspace{-48pt}
=\frac {G_F^2}{2\pi v_i} |V_{ud}|^2
|U_{ei}|^2 F(Z,E_e)
\frac{m_{^3 \rm He}}{m_{^3\rm H}}
E_e p_e A^h_i (\bar f^2+
3 \bar g^2), \nonumber
\label{eqC4}
\eeq
with $V_{ud}$ the up-down quark element of the CKM matrix, the $U_{ei}$ are the neutrino mixing matrix elements, and $F(Z,E_e)$ the Fermi Coulomb correction for 
the electron-$^3$He system.  The $\bar f$  and $\bar g$ are the nuclear form factors for Fermi and 
Gamow-Teller transitions, and the neutrino helicity-dependent factor is $A^\pm_i = 1\mp \beta_i$,
where $\beta_i=v_i/c$. 

    The total ITBD rate is given by $\sigma^h_i v_i$ integrated over the distribution (\ref{f0}) of neutrinos and summed over mass states $i$.   For Dirac neutrinos with spin rotated by $\theta_i$, both negative and positive helicity states, weighted by $\frac12(1\mp\cos\theta_i)$, contribute and
yield the neutrino dependence in the rate,
\beq
A_{\rm eff,D} &=&   \sum_{i,h=\pm}  |U_{ei}|^2\langle A_i^h\rangle_T = 
1 + \sum_i |U_{ei}|^2 \langle \beta_i\cos \theta_i \rangle_T.  \nonumber\\
\label{eqC5}
\eeq
The subscript $T$ includes the thermal average over the distribution (\ref{f0}) as well as the average
of the  spin rotation over the neutrino's history.

   Majorana neutrinos, as noted, have no diagonal magnetic moments and cannot flip spin in a slowly varying magnetic field, so that 
$\langle \cos\theta\rangle =1$.  Since the ITBD measures both Majorana neutrinos and antineutrinos, 
\beq
A_{\rm eff,M} &=& \big(1 + \sum_i |U_{ei}|^2\langle \beta_i \rangle_T\big) \nonumber \\ && \hspace{12pt}
+ \big(1 - \sum_i |U_{ei}|^2\langle \beta_i \rangle_T\big)  = 2, 
\label{eqC6}
\eeq
independent of the neutrino masses, and spin rotation by cosmic gravitational fluctuations \cite{grav}.

\begin{figure}[t]
\includegraphics*[width=\linewidth]{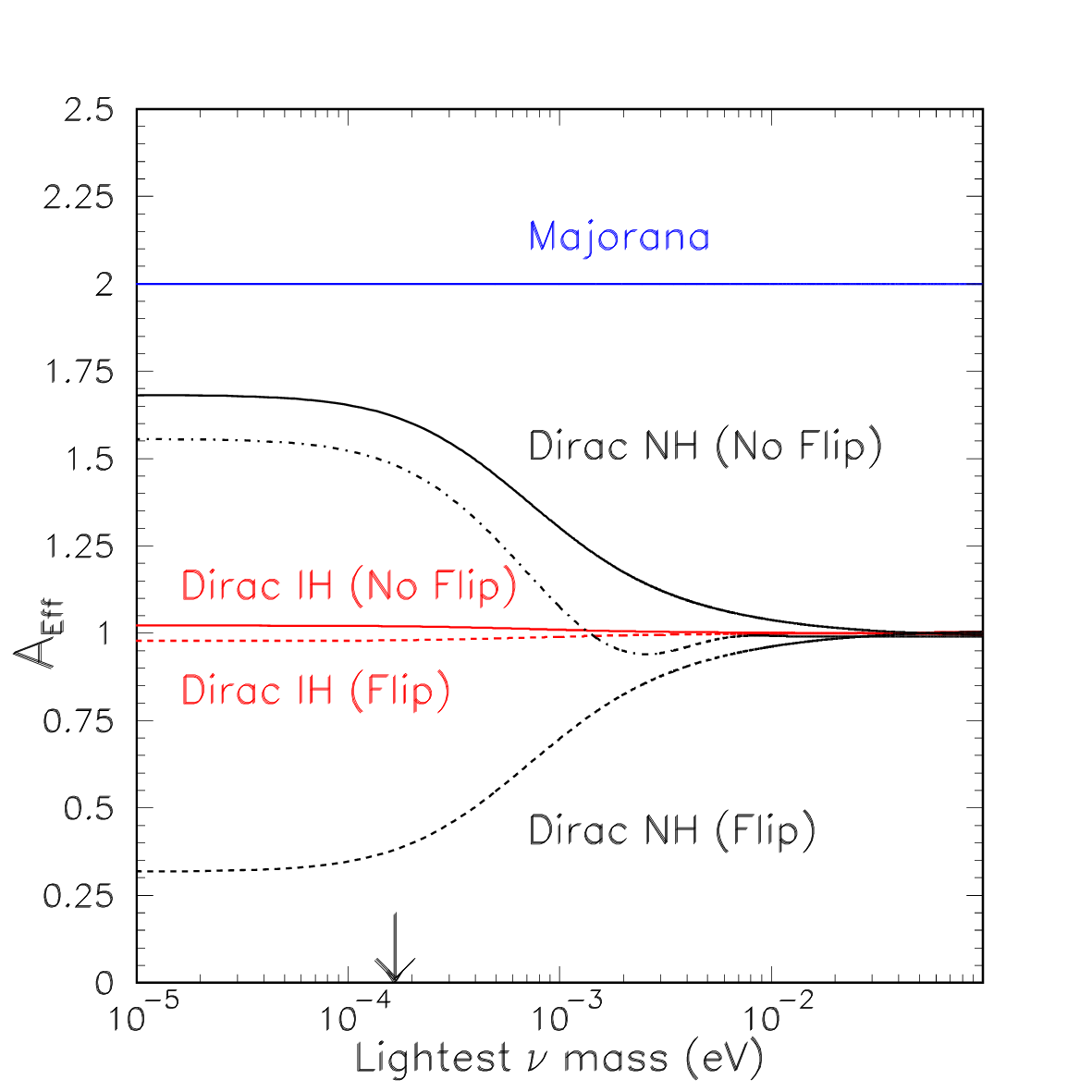}
\caption{
The coefficient $A_{\rm eff}$ vs. mass of the
lightest neutrino for Dirac and Majorana neutrinos, for the normal (NH) and inverted (IH) hierarchies.
  The dashed curves show
the extreme case of complete helicity flip from left to right handed.   The dash-dot curve shows the result for Dirac NH neutrinos with $\langle\theta^2\rangle$ given by the Milky Way estimate (\ref{galrot}), with  $B = 10 \,\mu$G, $\Lambda_g = 1\,$kpc, and 
$\mu_\nu =  5\times 10^{-14} \mu_B$.
The present neutrino temperature, $T_{\nu0}$ (arrow), demarcates the transition of the lightest neutrino from relativistic to non-relativistic.
}
\label{figV4}
\end{figure}

  Figure~\ref{figV4} shows the $A_{{\rm eff}}$ as a function of the mass of the lightest neutrino,
for both Dirac and Majorana neutrinos
with normal and inverted mass hierarchies.   For neutrinos maintaining their original helicity 
($\theta_i = 0$), the  $A_{\rm eff}$ are the solid curves.
 As the mass of the lightest neutrino approaches
zero, $A_{{\rm eff, D}}$ approaches $1 + |U_{e1}|^2 = 1.6794$ 
in the normal and $1 + |U_{e3}|^2 = 1.0216$ in the 
inverted hierarchy.   When the lightest neutrino mass rises and all neutrinos 
become nonrelativistic, $A_{{\rm eff,D}} =  1 +(7\pi^4T_{\nu0}/180\zeta(3)) \sum_i |U_{ei}|^2 /m_i$
eventually approaches unity independent of the mass hierarchy;  
 $A_{{\rm eff}}$ is always larger for Majorana than Dirac
neutrinos, independent of the mass hierarchy and the mass of the lightest
neutrino.

   The dashed curves in Fig.~\ref{figV4} show the dependence of $A_{\rm eff,D}$ on 
the lightest neutrino mass for complete helicity flip, 
$\theta_i = \pi$.  For partial spin rotation, $A_{\rm eff,D}$ lies
between the solid and dashed curves.    
When $\theta_i=\pi/2$, the amplitudes to be left and right handed are equal and $A_{\rm eff,D}=1$.  
To illustrate the qualitative dependence of the 
helicity-flip probability on $\mu_\nu$ in Fig.~\ref{figV4}, we show $A_{\rm eff,D}$ for 
Dirac neutrinos passing through the Milky Way as the dash-dot curve, calculated from Eq.~(\ref{galrot}) for small angle bending with $B_g = 10\,\mu$G and $\Lambda_g$= 1 kpc,  and 
with $\mu_\nu = 5\times10^{-14} \mu_B$, two orders of magnitude 
smaller than the magnetic moment XENON1T would need to explain their 
event excess. {The value of $\mu_\nu = 5 \times 10^{-14} \mu_B$ 
is also below the
upper bound derived from the analysis of solar neutrino 
data~\cite{Miranda04a,Miranda04b}, and is consistent with
the upper bound deduced from the stellar energy loss~\cite{Raffelt}.}  
If the magnetic moment 
of normal hierarchy Dirac neutrinos is of order that suggested
by XENON1T,  then for the characteristic parameters assumed for cosmic 
or galactic magnetic fields the neutrino spin rotations would no 
longer be small; the mean $\cos\theta$ would decrease $A_{\rm eff,D}$ to essentially unity, with a 
concomitant decrease in the ITBD detection rate.  A magnetic moment 
of the standard model prediction of Eq.~(\ref{numag}) 
would affect $A_{\rm eff,D}$ 
insignificantly. {In contrast, a value of 
$\mu_\nu = 10^{-14} \mu_B$,  the naturalness
upper bound obtained from an EFT analysis~\cite{bell-Dirac,bell}, would have a 
significant effect on $A_{\rm eff,D}$.}

  {Figure 1 illustrates how measurements of the rate of
relic neutrinos can distinguish Dirac from Majorana neutrinos, with an accuracy that will improve 
as knowledge of the correct hierarchy as well as the lightest 
mass come into sharper focus.}
 As the dash-dot curve indicates, the interesting regime is of bending not too small to be indistinguishable and not so large that all spins have comparable probability of being left and right handed.   This regime is characterized by a falloff in  $A_{\rm eff,D}$ and the ITBD detection rate with 
increasing light neutrino mass.    Unfortunately, it becomes increasingly difficult to resolve the relic neutrino events
from the tritium beta decay background for smaller neutrino mass.
 Inventing novel techniques to probe
the region of interest shown in Fig.~\ref{figV4} remains a challenge. 

    In conclusion, investigating the implications of a possible large
neutrino magnetic moment beyond that in the standard model on the helicities
of relic neutrinos as they propagate through the cosmic and galactic magnetic fields, we find
significant helicity modifications even if $\mu_\nu$ is two orders of magnitude smaller than that suggested
by the XENON1T result.    The present estimates of neutrino spin rotation can be sharpened by using detailed maps as well as numerical simulations of the astrophysical magnetic fields, e.g.,~\cite{bandj,han2018,han2019,imagine}.    In addition, the spin rotation of MeV energy neutrinos from the diffuse
supernova background \cite{dsbn} as well as from neutron stars \cite{fujikawa} is also potentially detectable, although using different experimental techniques than for relic neutrinos (e.g., with the Gd-doped 
Super-K detector and the inverse beta decay reaction) \cite{SN}.   

This research was supported in part by NSF Grant PHY18-22502.  We thank Michael Turner, and Tanmay Vachaspati for helpful discussions.


\begin{thebibliography}{99}
\bibitem{masses} The best-value fits to the mass squared differences, $\Delta m^2_{ij} \equiv m^2_i - m^2_j$, are \cite{esteban}
$\Delta m^2_{21}  =  (7.50 + 0.19 - 0.17) \times 10^{-5} {\rm eV}^2, 
\Delta m^2_{31,N}  =  (2.524 + 0.039 - 0.040)\times 10^{-3} {\rm eV}^2, 
\Delta m^2_{31,I}  =  (-2.514 + 0.038 - 0.041)\times 10^{-3} {\rm eV}^2$.
\bibitem{SN} Supernova and neutron star neutrinos are rotated by $\sim \mu_\nu B R/c$, in magnetic fields upwards of $10^{12}$ G, with $R$ the distance over which the magnetic fields are large.  For magnetic fields in neutron stars and supernovae reasonably coherent over the regime of high field then $ \theta^2 \sim 2\times10^{27} \left(R/10\,{\rm km}\right)^2 \left(B/10^{12}{\rm G}\right)^2 \left(\mu_\nu/\mu_B\right)^2$, 
comparable in scale to the rotation of relic neutrinos in galaxies.       
Characteristically, $B$ scales as $R^{-2}$ in core collapse supernovae, so that the rotation would be largest for neutrinos from neutron stars, during formation or later cooling.   The preponderance of diffuse background supernova and neutron star neutrinos are produced after redshift $z_s$ at most of order 10; the rotation of these neutrinos in cosmic magnetic fields, given by  Eq.~(\ref{inter}) using the integral in Eq.~(\ref{mde}) with $1/a_{eq} \to 1+z_s$, is negligible in comparison.
\end{thebibliography}
\end{document}